\documentstyle[12pt]{article}

\begin{document}
\newcommand{\balpha}{{\mbox{\boldmath$\alpha$}}}
\newcommand{\bmu}{{\mbox{\boldmath$\mu$}}}
\newcommand{\pprime}{{p^{\prime}}^0_1}
\newcommand{\pp}{{p^0_1}}
\newcommand{\vare}{\varepsilon}
\newcommand{\g}{\gamma}
\newcommand{\lbr}{\langle}
\newcommand{\rbr}{\rangle}
\newcommand{\bfx}{{\bf x}}
\newcommand{\bfy}{{\bf y}}
\newcommand{\bfz}{{\bf z}}
\newcommand{\intinf}{\int_{-\infty}^{\infty}}
%----------------------------------------------
\newcommand{\electronline}{\line(0,1){100}} %
\newcommand{\photarctop}{ %
  \multiput(0,0)(10,10){3}{\oval(10,10)[tl]}
  \multiput(0,10)(10,10){2}{\oval(10,10)[br]}
  \multiput(40,0)(-10,10){3}{\oval(10,10)[tr]}
  \multiput(40,10)(-10,10){2}{\oval(10,10)[bl]}
}
\newcommand{\pphotarctop}{ %
  \multiput(10,0)(10,10){2}{\oval(10,10)[tl]}
  \multiput(0,0)(10,10){3}{\oval(10,10)[br]}
  \multiput(0,40)(10,-10){3}{\oval(10,10)[tr]}
  \multiput(10,40)(10,-10){2}{\oval(10,10)[bl]}
}
\newcommand{\photonlineleft}{
 \multiput(0,0)(-16,0){3}{\oval(8,8)[b]}
 \multiput(-8,0)(-16,0){2}{\oval(8,8)[t]}}
\newcommand{\photonlineright}{
\multiput(0,0)(16,0){3}{\oval(8,8)[b]}
\multiput(8,0)(16,0){2}{\oval(8,8)[t]}}

\newcommand{\pphotarcbottom}{ %
  \multiput(0,0)(-10,10){3}{\oval(10,10)[bl]}
  \multiput(-10,0)(-10,10){2}{\oval(10,10)[tr]}
  \multiput(-10,40)(-10,-10){2}{\oval(10,10)[br]}
  \multiput(0,40)(-10,-10){3}{\oval(10,10)[tl]}
}
\newcommand{\ppphotarctop}{ %
   \multiput(0,0)(10,10){6}{\oval(10,10)[tl]}
   \multiput(0,10)(10,10){5}{\oval(10,10)[br]}
   \multiput(100,0)(-10,10){6}{\oval(10,10)[tr]}
   \multiput(100,10)(-10,10){5}{\oval(10,10)[bl]}
}
\newcommand{\loopbottom}{ %
  \put(0,0){\oval(5,5)[bl]}
  \multiput(0,-5)(0,-10){3}{\oval(5,5)[r]}
  \multiput(0,-10)(0,-10){2}{\oval(5,5)[l]}
  \put(0,-30){\oval(5,5)[tl]}
  \put(-30,-30){\electronline}
}
\newcommand{\ppphotwithlooptop}{ %
  \multiput(-1,0)(10,10){4}{\oval(10,10)[tl]}
  \multiput(-1,10)(10,10){4}{\oval(10,10)[br]}
  \multiput(101,0)(-10,10){4}{\oval(10,10)[tr]}
  \multiput(101,10)(-10,10){4}{\oval(10,10)[bl]}
  \put(50,37){\circle{30}}
}
%%%%%%%%%%%%%%%%%%%%%%%%%%%%%%%%%%%%%%%%%%

\begin{center}
{\Large Two-electron self-energy contribution to the ground state energy of
heliumlike ions}
\end{center}

\centerline{V. A. Yerokhin$^{1}$, A. N. Artemyev$^{2}$, and V. M. Shabaev$^{2}$}
\begin{center}
$^{1}${\it Institute for High Performance Computing and Data
Bases, Fontanca 118, St.Petersburg 198005, Russia}\\
{\it e-mail: yerokhin@snoopy.phys.spbu.ru}\\
$^{2}${\it Department of Physics, St.Petersburg State University,
 Oulianovskaya 1, Petrodvorets, St.Petersburg 198904, Russia}
\end{center}

\begin{abstract}
The two-electron self-energy
contribution to the ground state energy of heliumlike ions
is calculated
both for a
point nucleus and an extended nucleus in a wide interval of $Z$.
 All the
two-electron contributions are compiled to obtain most accurate values for the
two-electron part of the ground state energy of heliumlike ions in the
range $Z=20-100$. The theoretical value of the ground state
energy of $^{238}U^{90+}$, based on currently available theory,
is evaluated to be $-261382.9(8)$ eV, without higher order one-electron QED
corrections.
\end{abstract}

\centerline{PACS number(s): 31.30.Jv, 31.10.+z}

\section{Introduction}
The recent progress in heavy-ion spectroscopy provides
good perspectives for testing the quantum electrodynamics in a region of
strong electric field. In \cite{Marrs,Stohlker} the two-electron
contribution to the ground-state energy of some heliumlike ions was
measured directly by comparing the ionization energies of  heliumlike
and hydrogenlike ions. In such an
experiment the dominating one-electron contributions are
completely eliminated. Though the accuracy of the experimental results is
not high enough at present, it is expected \cite{Stohlker} that the
experimental accuracy will be improved by up to an order of magnitude
in the near future. This will provide
testing the QED effects in the second
order in $\alpha$.

In this paper we calculate  the ground state
two-electron self-energy correction in the second order in $\alpha$ in the
range $Z=20-100$. Calculations of this correction
were previously performed for some ions
for the case of a point nucleus by Yerokhin and Shabaev
\cite{Yerokhin95} and for an extended nucleus by Persson {\it et al.}
\cite{Persson96, Persson97}. Contrary our previous calculation of this
correction, the full-covariant scheme, based on
an expansion of the Dirac-Coulomb propagator in terms of interactions with
the external potential \cite{Snyderman,Schneider_PhD},
is used in the present work.  This technique was
already applied by the authors to calculate the self-energy correction
to the hyperfine splitting in hydrogenlike and lithiumlike ions
\cite{Yerokhin97,Shabaev97}.

The paper is organized as follows. In the  Sec. 2 we give a brief
outline of the calculation of the two-electron self-energy contribution. In
the Sec. 3 we summarize all the two-electron contributions to the
ground state energy of heliumlike ions.
 The relativistic units
($\hbar = c =1$) are used in the paper.

%%%%%%%%%%%%%%%%%%%%%%%%%%%%%%%%%%%%%%%%%%%%%%%%%%%%%%%%%%%%%%%%%%%%%%%%%%
\section{Self-energy contribution}

The two-electron self-energy contribution is represented by the Feynman
diagrams in Fig.1. The formal expressions for these diagrams can easily be
derived by the two-time Green function method \cite{Shabaev90}.
Such a derivation was discussed in detail in \cite {Yerokhin95}.
  The
diagrams in Fig.1a are conveniently divided into irreducible and reducible
parts. The reducible part is the one in which the intermediate state
energy (between the self-energy loop and the electron-electron interaction line)
coincides with the initial state energy. The irreducible part is the
remaining one.
The contribution of the irreducible part can be written in the same form
as the first order self-energy
\begin{eqnarray} \label{1}
\Delta E_{\rm irred} = 2\Bigl[ \lbr \xi |\Sigma_{\rm R}(\vare_a)|a\rbr
 + \lbr a|\Sigma_{\rm R}(\vare_a)|\xi \rbr \Bigr]\, ,
\end{eqnarray}
where $\Sigma_{\rm R}(\vare)$ is the regularized self-energy operator, $\vare_a$
is the energy of the initial state $a$, and $|\xi \rbr$ is a perturbed wave
function defined by
\begin{eqnarray} \label{2}
 |\xi \rbr =
 \sum_{\vare_n \neq \vare_a}
 \frac{|n\rbr \left[ \lbr nb| I(0)|ab\rbr - \lbr nb|I(0)|ba\rbr \right]}
  {\vare_a - \vare_n} \, .
\end{eqnarray}
Here $I(\omega )$ is the operator of the electron-electron interaction. The
calculation of the irreducible part is carried out using the scheme suggested
by Snyderman \cite{Snyderman} for the first order self-energy contribution.

The reducible part is grouped with the vertex part (Fig.1b).
For the sum of these terms the following formal expression is
obtained
\begin{eqnarray} \label{3}
\Delta E_{\rm vr} & = &{2\alpha} ^2 \sum_P (-1)^P \frac{i}{2\pi}
  \intinf d\omega \int  d\bfx d\bfy d\bfz
    \frac{e^{i|\omega ||\bfx -\bfy |}}{|\bfx -\bfy |}
 \nonumber \\ && \times
    \Biggl[ \psi ^{\dag}_{Pa}(\bfx ) \alpha _{\nu}
    \int d\bfz _1 \psi ^{\dag}_{Pb}(\bfz _1)
  \frac{\alpha _{\mu}}{|\bfz -\bfz _1|} \psi _{b}(\bfz _1)
 \nonumber \\ && \times
   G(\vare_a-\omega ,\bfx ,\bfz ) \alpha ^{\mu}
   G(\vare_a-\omega ,\bfz ,\bfy )
 \alpha ^{\nu} \psi _{a}(\bfy )
 \nonumber \\
&& \mbox{} - \lbr PaPb|\frac{1-\balpha_1 \balpha_2}{r_{12}} |ab\rbr
 \nonumber \\ && \times
 \psi ^{\dag}_{a}(\bfx ) \alpha _{\nu}
   G(\vare_a-\omega ,\bfx ,\bfz )
   G(\vare_a-\omega ,\bfz ,\bfy )
    \alpha ^{\nu} \psi _{a}(\bfy )\Biggr] \, .
\end{eqnarray}
Here the first term corresponds to the vertex part, and the second one
corresponds
to the reducible part. $G(\vare ,\bfx ,\bfz )$ is the Coulomb Green
function, $\alpha$ is the fine structure constant,
$\alpha^{\mu}=(1,\balpha)$,
 $\balpha$ are the Dirac
matrices, $a$ and $b$ are the $1s$ states with spin projection
$m=\pm \frac12$, and $P$ is the permutation operator.

According to the Ward identity the counterterms for the vertex and
reducible parts cancel each other, and, so, the sum of these terms
regularized in the same covariant way is ultraviolet finite. To cancel the
ultraviolet divergences analytically we divide $\Delta E_{\rm vr}$ into two
parts $\Delta E_{\rm vr} = \Delta E_{\rm vr}^{(0)}+ \Delta E_{\rm vr}^
{\rm many}$.
The first term  is $\Delta E_{\rm vr}$ with both
the bound electron propagators replaced by the free propagators.
It does not contain
the Coulomb Green functions and can be evaluated in the momentum
representation, where all the ultraviolet divergences are explicitly
cancelled using a standard covariant regularization procedure. The remainder
$\Delta E_{\rm vr}^{\rm many}$ does not contain ultraviolet divergent terms and
is calculated in the coordinate space.
The infrared divergent terms are handled
introducing a small photon mass $\mu$. After these
 terms are separated and cancelled
analytically the limit $\mu \to 0$ is taken.

In practice the calculation of the self-energy contribution is made using
the shell model of the nuclear charge distribution. Since the finite nuclear
size effect is small enough even for high $Z$ (it constitutes about
1.5 percent for
uranium), an error due to incompleteness of such a model is negligible. The
Green function for the case of the shell nucleus in the form presented in
\cite{Gyulassy} is used in the calculation. To calculate the part of
$\Delta E_{\rm irred}$ with two and more external potentials, we
subtract from the Coulomb-Dirac Green function the  first two terms
of its potential expansion numerically. To obtain the second term of
the expansion it is necessary to evaluate a derivative of the Coulomb Green
function with respect to $Z$ at the point $Z=0$. We handle it using
some algorithms suggested in \cite{Manakov}.

The numerical evaluation of $\Delta E_{\rm vr}^{\rm many}$ is the most time
consuming part of the calculation. The energy integration is carried out
using the Gaussian quadratures after rotating  the integration contour into
imaginary axis.  To achieve a desirable
precision it is sufficient to calculate 12-15 terms of the partial wave
expansion. The remainder is evaluated by fitting the partial wave
contributions to a polynomial in $\frac1l$. A contribution
arising from the intermediate electron states which are of the same energy
as the initial state is calculated separately using the
B-spline method for the Dirac equation \cite{Johnson}. The same method is
used for the numerical evaluation of the perturbed wave function $|\xi
\rbr $ in equation (1).

Table 1 gives the numerical results for the two-electron self-energy
contribution to the ground state energy
of heliumlike ions expressed in terms of
the function $F(\alpha Z)$ defined by
\begin{eqnarray}\label{4}
\Delta E = \alpha^2 (\alpha Z)^3 F(\alpha Z)\,mc^{2}
\end{eqnarray}
To the lowest order in $\alpha Z$, $F=1.346\ln{Z}-5.251$ (see
\cite{Yerokhin95} and references therein).
The results for a point nucleus and an extended nucleus are listed in the third
and fourth columns of the table, respectively. In the second column
the values of the root-mean-square (rms) nuclear charge radii
used in the calculation
 are given
\cite{Fricke,Johnson85}.
In the fifth column the results for an extended nucleus expressed in eV
are given to be compared with the ones of Persson {\it et al.}
\cite{Persson96} listed in the last column of the table.
A comparison of the present results
for a point nucleus
with the ones
from \cite{Yerokhin95} finds some discrepancy
for the contribution which corresponds to the Breit part of the
electron-electron interaction.
This discrepancy results from a small spurious term arising in the non-covariant
regularization procedure used in \cite{Yerokhin95}.

%%%%%%%%%%%%%%%%%%%%%%%%%%%%%%%%%%%%%%%%%%%%%%%%%%%%%%%%%%%%%%%%%%%%%%%%%%
\section{The two-electron part of the ground state energy}

In the Table 2 we summarize all the two-electron contributions to the ground
state energy of heliumlike ions. In the second column
of the table the energy contribution due to one-photon exchange is given.
Its calculation is carried out for the Fermi model of the nuclear charge
distribution
\begin{eqnarray}
\rho(r) = \frac{N}{1+\exp{((r-c)/a)}}\,
\end{eqnarray}
with the rms charge radii listed in the Table 1.
Following to \cite {Fricke}, the parameter $a$ is chosen to be $a =
 \frac{2.3}{4\ln 3}\,$ fm.
The parameters $c$ and $N$, with a good precision,
are given by
(see, e.g.,  \cite{Shabaev93})
\begin{eqnarray}
&c = \frac1{\sqrt{3}}\left[ \Bigl (4\pi^4a^4-10\lbr r^2\rbr \pi^2a^2+
\frac{25}{4}
\lbr r^2\rbr^2\Bigr )^{\frac12}
 -5\pi^2a^2+ \frac52\lbr r^2\rbr
\right]^{\frac12} \,,  \\
&N=\frac{3}{4\pi c^{3}}\Bigl(1+\frac{\pi^{2}a^{2}}{c^{2}}\Bigr)^{-1}\,.
\end{eqnarray}
Except for $Z$=83, 92,
the uncertainty of this correction
is obtained by a one percent variation of the rms radii.
In the case $Z$=92 ($\langle r^{2}\rangle^{1/2}=5.860(2)$ fm
 \cite{Zumbro84}), the uncertainty of this correction
is estimated by taking the difference between the corrections obtained with
the Fermi model and the homogeneously charged sphere model of the same
rms radius. For $Z=83$, the uncertainty comes from both a
variation of the rms radius by 0.020\,fm (it corresponds to a discrepancy
between the measured rms values \cite{Fricke})
and the
difference between the Fermi model and the homogeneously charged sphere model.

The energy contribution due to two-photon exchange is divided into two parts.
The first one ("non-QED contribution")
includes the non-relativistic contribution and the lowest order
($\sim (\alpha Z)^{2}$) relativistic
correction, which can be derived from the Breit equation.
This is given by the first two terms
in the $\alpha Z$-expansion \cite{Sanders,Palchikov,Drake}
\begin{eqnarray}
\Delta E_{\rm non-QED} = \alpha^{2}[-0.15766638 - 0.6356 (\alpha Z)^2]mc^{2}
\end{eqnarray}
and is presented in the third column of the Table 2.  The second part
which we refer to as the "QED
contribution" is the residual and is given in the fourth column of the
table. The data for the two-photon contribution for all $Z$, except for $Z=92$,
are taken from \cite{Blundell}, interpolation is made when it is
needed.  For $Z=92$ data from \cite{Lindgren} are taken.
In the fifth
column of the table the results of the present calculation of the
two-electron self-energy contribution are given. The two-electron vacuum
polarization contribution  taken from \cite{Artemyev}  is listed in the
sixth column.
In the seventh
column the "non-QED part" of
the energy correction due to exchange of three and more photons
is given.  This correction is evaluated by summing
the $Z^{-1}$ expansion terms for the ground state energy of heliumlike ions
beginning from $Z^{-3}$.  The coefficients
of such an expansion are taken to zeroth order in $\alpha Z$ from
\cite{Sanders} and to second order in $\alpha Z$ from \cite{Drake}.
The three and more photons QED correction has not yet been calculated.
We assume that the uncertainty due to omitting this correction is
of order of magnitude of the total second-order QED correction multiplied
by factor $Z^{-1}$. It is given in the eighth column of the table.
The two-electron nuclear recoil correction is estimated by reducing
the one-photon exchange contribution by the factor $(1-m/M)$.
Such an estimate corresponds to the non-relativistic treatment
of this effect and takes into account that the mass-polarization
 correction is negligible for the $(1s)^2$ state \cite{Drake}.
This correction and its uncertainty, which is taken to be
100\% for high $Z$, are included into the total two-electron contribution.
The two-electron
nuclear polarization effect is expected to be negligible for the ground
state of heliumlike ions.
In the last column the total two-electron part
of the ground state energy of heliumlike ions is given.

In the Table 3 our results are compared with the experimental data
\cite{Marrs, Stohlker} and the results of previous calculations
based on the unified method \cite{Drake}, the all-order relativistic many
body perturbation theory (RMBPT) \cite{Plante}, the multiconfiguration
Dirac Fock treatment \cite{Indelicato}, and RMBPT with the complete
treatment of the two-electron QED correction \cite{Persson96,Persson97}.
Data in the third column of the table are taken from
\cite{Persson96} for $Z=54, 92$ and from \cite{Persson97} for other $Z$.  The
one-electron contribution from \cite{Johnson85} is subtracted from the
total ionization energies presented in \cite{Plante, Drake} to obtain the
two-electron part.

In the Table 4 we present the theoretical contributions to the ground state
energy of $^{238}U^{90+}$,
based on currently available theory.  The uncertainty of the
one-electron Dirac-Coulomb value comes from the uncertainty
of the Rydberg constant (we use $hcR_{\infty}$=13.6056981(40) eV,
$\alpha$=1/137.0359895(61)).
The one-electron nuclear size correction for
the Fermi distribution with $\lbr r^2 \rbr
^{1/2} = 5.860 \,$fm gives $397.62(76)\,$ eV. The uncertainty of this correction
is estimated by taking the difference between the corrections obtained with
the Fermi model and the homogeneously charged sphere model of the same
rms radius \cite{Franosch}.
The nuclear recoil correction was  calculated to all orders in $\alpha
Z$ by Artemyev {\it et al.} \cite{Artemyev2}.
The uncertainty of this correction is chosen to include
a deviation from a point nucleus approximation used in
\cite{Artemyev2}.
The one-electron nuclear
polarization effect was evaluated  by Plunien and Soff \cite{Plunien}
and by Nefiodov {\it et al.} \cite{Nefiodov}.
The values of the first order
self-energy and vacuum polarization corrections are taken from
\cite{Mohr} and \cite{Persson93}, respectively.
 The two-electron corrections are quoted from the Table 2.
The higher order one-electron QED
corrections are omitted in this summary since they have not yet been calculated
completely. We expect they can contribute within several electron
volts.

\section*{Acknowledgments}
Valuable conversations with Thomas St\"ohlker are gratefully
acknowledged.
The research described in this publication was made possible in part by
Grant No. 95-02-05571a from the Russian Foundation for Basic
Research.

\newpage
\small
%%%%%%%%%%%%%%%%%%%%%%%%%%%%%%%%%%%%%%%%%%%%%%%%%%%%%%%%%%%%%%%%%%%%%%%%%%%
%
%       Table 1
%
%%%%%%%%%%%%%%%%%%
\begin{table}
 \caption{The two-electron self-energy contribution to the ground state
 energy of
 heliumlike ions. The function $F(\alpha Z)$ is defined by equation (4).}
\vspace{1cm}
\begin{tabular}{|r|c|c|c|c|c|}
\hline
$Z$  & $\lbr r^2\rbr^{1/2}$[fm]& $F$ [point.nucl.] &$F$ [ext.nucl.]&
$\Delta E$ [eV] (this work)
 & $\Delta E$ [eV] (Ref. \cite{Persson96}) \\ \hline

20       &                &  -1.7134(3)  &                 & -0.1449 &      \\
30       &    3.928       &  -1.3889(3)  &      -1.3887(3) & -0.3965(1) &   \\
32       &    4.072       &  -1.3448(3)  &      -1.3446(3) & -0.4659(1) &-0.5\\
40       &    4.270       &  -1.2115(3)  &      -1.2112(3) & -0.8197(2) &   \\
50       &    4.655       &  -1.1140(3)  &      -1.1134(3) & -1.4717(4) &   \\
54       &    4.787       &  -1.0907(3)  &      -1.0898(3) & -1.8146(5) &-1.8\\
60       &    4.914       &  -1.0692(3)  &      -1.0679(3) & -2.4391(7)&     \\
66       &    5.224       &  -1.0625(3)  &      -1.0603(3) & -3.223(1)  &-3.2\\
70       &    5.317       &  -1.0657(3)  &      -1.0629(3) & -3.855(1)  &    \\
74       &    5.373       &  -1.0752(3)  &      -1.0712(3) & -4.590(1)  &-4.6\\
80       &    5.467       &  -1.1014(3)  &      -1.0951(3) & -5.929(2)  &   \\
83       &    5.533       &  -1.1205(3)  &      -1.1124(3) & -6.726(2)  &-6.7\\
90       &    5.645       &  -1.1830(3)  &      -1.1682(3) & -9.005(2)  &    \\
92       &    5.860       &  -1.2062(4)  &      -1.1878(4) & -9.780(3)  &-9.7\\
100      &    5.886       &  -1.3307(6)  &      -1.2929(6) &-13.671(6)  &    \\
\hline
\end{tabular}
\end{table}
\scriptsize
%%%%%%%%%%%%%%%%%%%%%%%%%%%%%%%%%%%%%%%%%%%%%%%%%%%%%%%%%%%%%%%%%%%%%%%%%%%
%
%       Table 2
%
%%%%%%%%%%%%%%%%%%
\begin{table}
 \caption{Varios components of the two-electron contribution to the
 ground-state energy of helium-like ions (in eV).}
 \vspace{1cm}
\begin{tabular}{|r|l|c|c|c|l|c|c|l|}         \hline
$Z$ &1-ph. exch.&2-ph. exch.&
 2-ph. exch.& Self   & Vac.
&
$\geq 3$ ph.& $\geq 3$ ph.&
Total contr. \\
    &   &non-QED   &
         QED    & energy &
pol.&
     non-QED   &    QED      &
       \\ \hline
20 &\  345.76    & -4.66 &
 \  0.01   & -0.15   & 0.01  &
 0.03 & $\pm$0.01&
 \
341.00(1) \\

30 &\  529.42    & -5.12 &
 \  0.04   & -0.40   & 0.04  &
 0.03 & $\pm$0.01&
 \
524.00(1) \\

32 &\  567.61    & -5.23 &
 \  0.04   & -0.47   & 0.04  &
 0.03 & $\pm$0.01&
 \
562.02(1) \\

40 &\  726.64    & -5.76 &
\  0.07   & -0.82   & 0.09  &
0.03 & $\pm$0.02&
\
720.24(2) \\

50 &\  943.09    & -6.59 &
\  0.10   & -1.47   & 0.19  &
0.04 & $\pm$0.02&
\
935.35(2) \\

54 &  1036.56    & -6.98 &
 \  0.10   & -1.82   & 0.26  &
 0.04 & $\pm$0.03&
1028.16(3) \\

60 &  1185.73(1) & -7.61 &
\  0.09   & -2.44   & 0.38  &
0.04 & $\pm$0.03&
1176.19(3) \\

66 &  1347.45(1) & -8.30 &
 \  0.06   & -3.22   & 0.56  &
 0.05 & $\pm$0.04&
1336.58(4) \\

70 &  1463.43(1) & -8.80 &
\  0.02   & -3.86   & 0.71  &
0.05 & $\pm$0.05&
1451.55(5) \\

74 &  1586.93(2) & -9.33 &
  -0.04(1)& -4.59   & 0.91  &
  0.05 & $\pm$0.05&
1573.92(6) \\

80 &  1788.43(3) &-10.19 &
   -0.19   & -5.93   & 1.30  &
0.06 & $\pm$0.06&
1773.47(7) \\

83 &  1897.56(1) &-10.64 &
   -0.30(1)& -6.73   & 1.55  &
0.06 & $\pm$0.07&
1881.50(7) \\

90 &  2178.25(7) &-11.75 &
  -0.65   & -9.01   & 2.34  &
  0.06 & $\pm$0.08&
2159.24(11) \\

92 &  2265.88(1) &-12.09 &
  -0.79   & -9.78   & 2.63  &
   0.06 & $\pm$0.09&
2245.92(9) \\

100&  2659.8(2)  &-13.50 &
   -1.58   &-13.67(1)& 4.25  &
   0.07 & $\pm$0.11&
2635.4(2)   \\ \hline
\end{tabular}
\end{table}
\normalsize
%%%%%%%%%%%%%%%%%%%%%%%%%%%%%%%%%%%%%%%%%%%%%%%%%%%%%%%%%%%%%%%%%%%%%%%%%%%
%
%       Table 3
%
%%%%%%%%%%%%%%%%%%
\begin{table}
 \caption{Two-electron contribution to the ground state energy of
 some helium-like ions (in eV).}
 \vspace{1cm}
\begin{tabular}{|c|c|c|c|c|c|c|}
\hline
$Z$ & Present work & Persson {\it et.al}        &Indelicato       &Plante
{\it et al.} & Drake      & Experiment\\
    &              &
\cite{Persson96,Persson97}&\cite{Indelicato}&\cite{Plante}   $^a$
&\cite{Drake}& \cite{Marrs,Stohlker}\\
\hline

32 &\, 562.02(1) &\, 562.02(10)  &\,562.1 &\,562.1 & \,562.1 &\,562.5(1.6)\\
54 &  1028.16(3) &  1028.2 \, \, \, \, & 1028.2 & 1028.4 &  1028.8 &
1027.2(3.5)\\
66 &  1336.58(4) &  1336.59(10)  & 1336.5 & 1337.2 &  1338.2 & 1341.6(4.3)\\
74 &  1573.92(6) &  1573.93(10)  & 1573.6 & 1574.8 &  1576.6 & 1568(15) \, \, \\
83 &  1881.50(7) &  1881.54(10)  & 1880.8 &        &  1886.3 & 1876(14) \, \, \\
92 &  2245.92(9) &  2246.0 \, \, \, \, &     &        &  2255.1 & \\\hline
\end{tabular}
\normalsize
$^a$ taken from \cite{Marrs}.
\end{table}
%%%%%%%%%%%%%%%%%%%%%%%%%%%%%%%%%%%%%%%%%%%%%%%%%%%%%%%%%%%%%%%%%%%%%%%%%%%
%
%       Table 4
%
%%%%%%%%%%%%%%%%%%
\begin{table}
\caption{Theoretical contributions to the ground state energy of
$^{238}U^{90+}$, without higher order one-electron QED corrections.}
\begin{center}
\begin{tabular}{|l|r|r|}\hline
Contribution          &   Value [eV]   & Uncertainty [eV]    \\ \hline
One-electron Dirac    &  -264559.97   & $\pm$ 0.08  \\
One-electron nuc. size&      397.62   & $\pm$ 0.76 \\
One-electron nuc. rec.&        1.02   &$\pm$0.10   \\
One-electron nuc. pol.&       -0.40   & $\pm$ 0.10 \\
1st order self-energy &      710.09   &   \\
1st order vac. pol.   &     -177.20   &   \\
1-ph. exchange        &     2265.88   & $\pm$ 0.01  \\
2-ph. exchange non-QED&      -12.09   &   \\
2-ph. exchange QED    &       -0.79   &  \\
Self-energy screening &       -9.78   &   \\
Vac. pol. screening   &        2.63   &   \\
$\geq$ 3-ph. non-QED  &        0.06   &  \\
$\geq$ 3-ph QED       &        0.00   & $\pm$ 0.09\\
Two-electron nuc. rec. &       -0.01   & $\pm$ 0.01\\ \hline
Total                 &  -261382.9 \,  & $\pm$ 0.8 \, \\ \hline
\end{tabular}
\end{center}
\end{table}

\newpage
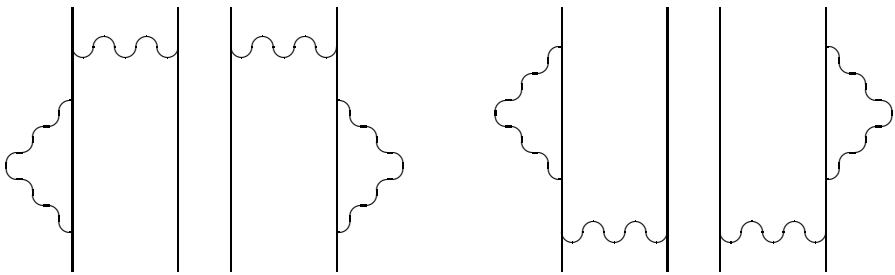
\begin{figure}
\caption{Two-electron self-energy
diagrams.}
  \begin{picture}(480,400)
    \put(250,30){
      \put(65,0){\electronline}
      \put(65,30){\pphotarctop}
      \put(61,50){\photonlineleft}
      \put(25,0){\electronline}
        }
    \put(230,10){b}

    \put(110,30){
      \put(40,0){\electronline}
      \put(40,30){\pphotarcbottom}
      \put(44,50){\photonlineright}
      \put(80,0){\electronline}
    }

   \put(125,250){
      \put(65,0){\electronline}
      \put(65,20){\pphotarctop}
      \put(61,85){\photonlineleft}
      \put(25,0){\electronline}
      }
   \put(230,220){a}

    \put(50,250){
      \put(40,0){\electronline}
      \put(40,20){\pphotarcbottom}
      \put(44,85){\photonlineright}
      \put(80,0){\electronline}
          }
   \put(310,250){
      \put(65,0){\electronline}
      \put(65,40){\pphotarctop}
      \put(61,15){\photonlineleft}
     \put(25,0){\electronline}
    }
   \put(235,250){
      \put(40,0){\electronline}
      \put(40,40){\pphotarcbottom}
      \put(44,15){\photonlineright}
      \put(80,0){\electronline}

    }
  \end{picture}
\end{figure}
%%%
%%%%%%%%%%
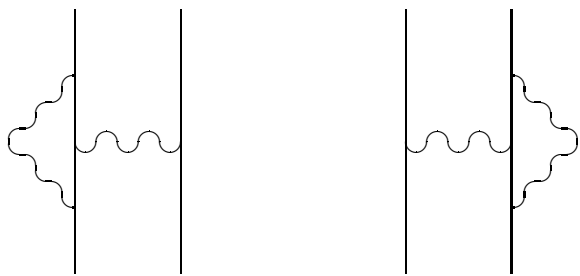
\end{document}